\documentclass[11pt]{article}

\input epsf.sty

\def\lromn#1{\uppercase\expandafter{\romannumeral#1}}

\usepackage{graphicx,amsmath,amssymb}
\def\lromn#1{\uppercase\expandafter{\romannumeral#1}}

\begin{document}

\vspace{2cm}
\begin{center}
\begin{Large}
{\bf Parity violating radiative emission of neutrino pair in 
heavy alkaline earth atoms of even isotopes
}

\end{Large}

\vspace{2cm}
\begin{large}
M. Yoshimura, 
 N. Sasao$^{\dagger}$, and S. Uetake

Center of Quantum Universe, Faculty of
Science, Okayama University \\
Tsushima-naka 3-1-1 Kita-ku Okayama
700-8530 Japan

$^{\dagger}$
Research Core for Extreme Quantum World,
Okayama University \\
Tsushima-naka 3-1-1 Kita-ku Okayama
700-8530 Japan \\
\end{large}
\end{center}

\vspace{4cm}

\begin{center}
\begin{Large}
{\bf ABSTRACT}
\end{Large}
\end{center}

Metastable excited states ${}^3P_2, {}^3P_0$ of heavy alkaline earth
atoms of even isotopes are studied for parity violating (PV) effects in
radiative emission of neutrino pair (RENP).
PV terms  arise from interference between two diagrams containing
neutrino pair emission of valence spin
current and nuclear  electroweak charge density proportional
to the number of neutrons in nucleus. 
This mechanism gives large PV effects, since
it does not suffer from the suppression of 1/(electron mass)
usually present for non-relativistic atomic electrons.
A controllable magnetic field is crucial to identify
RENP process by measuring PV observables.
Results of PV asymmetries under the magnetic
field reversal and the photon circular polarization reversal 
are presented for an example of Yb atom.

\vspace{4cm}

Key words

Neutrino mass,
Parity violation, Majorana particle,
Beyond the standard gauge theory

\newpage
\lromn1 
{\bf Introduction}

\vspace{0.3cm}
For an unambiguous test of the weak nature
of interaction it is crucial to directly
observe odd quantities under parity operation.
Parity violation in atomic transitions has been 
one of the key steps towards verification
of the neutral current structure in electron
interaction with nucleus.
Mixture of different parity states in heavy
atoms \cite{bouchiat 2} is caused by
Z-boson exchange interaction with nucleus and its existence
has been verified in atomic parity violation experiments 
\cite{bouchiat exp}, \cite{commins},
\cite{wieman}.

A hint of new physics beyond the standard gauge theory of $SU(3) \times SU(2) \times U(2)$
has been found in neutrino oscillation experiments, establishing
finite neutrino masses with mixing.
The first stage of oscillation experiments
has been able to determine two mass squared differences
and three mixing angles \cite{nu oscillation data}.
The next important steps are to determine 
(1) the mass difference
pattern, the normal vs inverted mass hierarchical pattern,
(2) the absolute neutrino mass scale or the smallest
neutrino mass, and (3) determination of the nature of mass terms,
Majorana or Dirac mass, along with their CP properties.
Besides the oscillation experiments nuclear targets
are main tools of ongoing experiments \cite{tritium},
\cite{nu0 beta}.

We proposed a new method towards a future neutrino
physics; the use of atoms.
Parity violation is important to the new proposed
process of macro-coherent radiative emission
of neutrino pair (RENP),
$|e \rangle \rightarrow |g\rangle + \gamma +\nu \nu$
from metastable atomic state $|e\rangle$
\cite{my-prd-07}, \cite{ptep overview},
to demonstrate that the weak interaction is involved,
thereby establishing experimental identification of RENP
under a possible presence of QED backgrounds.
We advanced a step forward towards this direction,
and studied PV effects in alkaline earth atoms
of odd isotopes \cite{ysu-pv-13}.
Alkaline earth atoms are excellent for the purpose of PV effects,
since two low lying metastable states of ${}^3P_2, {}^3P_0$
for the initial RENP $|e\rangle$ state
have different parity from the ground $|g\rangle$ state,
which is required for PV effects.
PV arises from interference between parity odd (PO) and
parity even (PE) amplitudes.
In the scheme of \cite{ysu-pv-13} hyperfine interaction with nucleus
of odd isotopes has been used in the PE amplitude.
In the present work we shall examine alkaline earth atoms
of even isotopes where hyperfine interaction is absent.

We rely on an external magnetic field for even isotopes to mix
$J=2,0$ state with $J=1$ state necessary for PE amplitude
of intermediate
transition ${}^3P_{2,0} \rightarrow {}^{\pm}P_1 \rightarrow {}^1S_0$
($ {}^{\pm}P_1$ is the mixture of ${}^3P_1$ and ${}^1P_1$ 
caused by  spin-orbit interaction).
The mixing amplitude by the magnetic field
is of order $\mu_B B \sim  50 \mu {\rm eV} B /{\rm Tesla}$
divided by energy difference of levels, to be
compared with hyperfine mixing of $O(\mu {\rm eV})$
\cite{ysu-pv-13}.
The advantage of the external magnetic field
in alkaline earth atoms of even isotopes
has been demonstrated in another context,
the clock transition
of Yb atom \cite{b-assited clock transition}.
It turns out that the required Coulomb interaction with nucleus
for RENP PE amplitude  gives rise to a large amplitude 
in accordance with discussion in \cite{ys-13}.
Thus, the magnetic field
application may also be important to 
achieve a large enhancement for alkaline earth atoms
of odd isotopes, but we shall discuss only even isotopes
in order to avoid unnecessary complications of the mechanism.
Another merit of the applied magnetic field in even isotopes
is its controllability of magnitudes and
direction in measurement of PV observables.
It should thus help much in identification of
RENP process in experiments.

In a series of theoretical papers
we developed and gradually refined a new, systematic
experimental method to probe the neutrino mass matrix
using RENP.
Following the initial idea \cite{my-prd-07},
we first discussed how to enhance otherwise small 
neutrino pair emission rates 
\cite{yst pra}, \cite{ptep overview},
and then how to extract neutrino parameters
from the photon energy spectrum \cite{dpsty-plb}, \cite{ptep overview}.
In the most recent work
we pointed out how to obtain a much larger RENP
rate \cite{ys-13} using a coherent neutrino
pair emission from nucleus where the zero-th
component of vector current operates 
much like the enhanced admixture of 
different parity states in atomic PV experiments.
Our experimental efforts towards RENP 
are briefly described in \cite{ptep overview}.
Clearly, investigation of PV effects
is the next important step in RENP.

The rest of this work is organized as follows.
In Section \lromn2 how PV observables
may arise in the standard electroweak theory
(with finite neutrino masses)
by listing all PO and PE pair emission
vertexes to the leading and the next sub-leading orders
of 1/mass.
Some technical details on the phase space integral
of neutrino pair variables (helcities and momenta) that have
a direct relevance to emergence of
parity odd quantities are relegated to Appendix A.
We then calculate in Section \lromn3 amplitudes of RENP,
emphasizing how the magnetic field dependence
is disentangled.
In Section \lromn4 RENP rates, both parity
conserving (PC) and PV, are calculated.
PC rates and PV asymmetries are
given in analytic forms using explicitly known elementary
functions: dependences on parameters of
the neutrino mass matrix elements are thereby clearly worked out.
We then illustrate results of numerical computations
on PV observables and its asymmetry under the magnetic field and the
photon circular polarization reversals,
taking the example of the Yb $J=2 \rightarrow 0$ transition.
(PV effects are found to vanish for ${}^3P_0$.)
The rates have an overall uncertain factor subject to
detailed numerical simulations dependent on experimental
conditions.
The spectral shape is however determined unambiguously
as function of neutrino parameters.
We are able to present spectral shapes and PV asymmetries
assuming a single  unknown parameter of smallest neutrino mass and
taking other parameters consistent with the present oscillation
data \cite{nu oscillation data}.
Rates related to PV are insensitive to Majorana CP phases,
but PV observables can measure the smallest mass,
and make distinction of normal and inverted hierarchical mass
pattern, and
distinction of Majorana and Dirac neutrino,
The rest of this work consists of summary and Appendices.

We are bound to calculate amplitudes using perturbation theory 
in non-relativistic quantum mechanics, hence
the time ordering in higher orders
of perturbation should be treated with care.
This gives rise to cancellations of
a few added contributions.

Throughout this work we use the natural unit
of $\hbar = c = 1$.

\vspace{0.5cm}
\lromn2 
{\bf Candidate search for parity odd and even amplitudes}

\vspace{0.3cm}
Typical RENP experiments use several lasers
for trigger and excitation.
For instance, two continuous wave (CW) lasers
of different frequencies $\omega_i, i=1,2$ where
$\omega_1 < \omega_2\,, \omega_1 +\omega_2 = \epsilon_{eg}$
and $\epsilon_{eg}$ is
the energy difference between the initial $|e\rangle$ state and 
the final $|g\rangle$ state,
are used as triggers in counter propagating directions (taken along z-axis),
while two excitation lasers of Raman type of frequencies,
$\omega_{p}\,,\omega_{s}$ with  $\omega_{p}-\omega_{s} = \epsilon_{eg}$
are irradiated in pulses.
Measured variables at the time of excitation pulse irradiation
are the number of events at each trigger frequency $\omega_1$.
By repeating measurements at different trigger frequency combinations, 
one obtains
the photon energy spectrum at different frequencies 
$\omega = \omega_1 $ accompanying the invisible neutrino pair.
If PV effects are large,
measurements of PV asymmetries help reject QED  backgrounds,
the largest being two-photon emission.

The macro-coherent three-body RENP process 
$|e \rangle \rightarrow |g\rangle + \gamma +\nu \nu$
conserves both the energy and the momentum,
giving continuous photon energy spectrum with thresholds.
Note that the spontaneous decay of dipole transition
from excited atoms conserves the energy alone, hence their spectrum 
is continuous despite of a single particle decay.
In RENP there are six photon energy thresholds at
$\omega_{ij} = \epsilon_{eg}/2 - (m_i+ m_j)^2/2\epsilon_{eg}$
with $ m_{ij}\,(i,j = 1,2,3)$ 
three neutrino masses of mass eigenstates. 
Decomposition into six different threshold regions 
is made possible by excellent energy resolution of 
trigger laser frequencies.

PV effects arise from interference of two RENP amplitudes
of parity even (PE) and parity odd (PO). 
Note that both rates arising from the squared
PO and the squared PE amplitudes give  PC rates.
There are two types of neutrino pair emission amplitudes with regard to
spatial behavior, $A_0 \nu_i^{\dagger} \nu_j$, 
and $\vec{A}\cdot\nu_i^{\dagger} \vec{\sigma}\nu_j  $,
where $A_0$ is atomic matrix element
relevant to the nuclear mono-pole current of
neutrino pair emission, and $\vec{A}$ is the one relevant to
the spin current from valence electron.
Each of $A_{\alpha}, \alpha = 0, 1,2,3$ contains product of 
E1 matrix elements, couplings
and energy denominators in perturbation theory.
We use two component notation for electron operators 
in the neutrino emission vertex of $A_{\alpha}$,
following 
the $\gamma_5$-diagonal representation of \cite{my-prd-07}.
Relevant leading terms for PO and PE terms 
for pair emission of mass eigenstates $ij$ are given by
\begin{eqnarray}
&&
A_0 \propto e^{\dagger}\left(b_{ij}
+ \delta_{ij}
2 \sin^2\theta_w \vec{\sigma}\cdot \frac{\vec{p}}{m_e}
 + O(\frac{1}{m_e^2})\right) e
+\delta_{ij}  j_q^0
\,,\hspace{0.5cm}
j_q^0 = -\frac{1}{2} j_n^0 + \frac{1}{2}(1-4\sin^2 \theta_w) j_p^0
\,,
\\ &&
\vec{A} \propto e^{\dagger}
\left(a_{ij} \vec{\sigma}  
+\delta_{ij} 2 \sin^2\theta_w \frac{1}{m_e} (\vec{p} - i \vec{\sigma} \times \vec{p}) + O(\frac{1}{m_e^2}) \right) e
\,,
\\ &&
a_{ij} = - U_{ei}^* U_{ej} + \frac{1}{2} \delta_{ij}
\,, \hspace{0.5cm}
b_{ij} = U_{ei}^* U_{ej} - \frac{1}{2} \delta_{ij}
(1 - 4\sin^2 \theta_w)
\,,
\end{eqnarray}
where necessary neutrino mixing matrix elements
$U_{ei}$  have been determined experimentally \cite{nu oscillation data}
whose values we use
in our following analysis.
The weak mixing angle is determined experimentally;
$\sin^2 \theta_w \sim 0.238$.
The term $j_q^0$ is the nuclear mono-pole current contribution
which gives rise to coherently added constituent numbers \cite{ys-13}.
We disregarded terms of orders
of $1/m_e^2$ and $1/m_N$,

In order to calculate  parity conserving (PC) and 
parity violating (PV) rates,
added amplitudes are squared, and one proceeds to
calculate summation over neutrino helicities and
momenta, since neutrino
variables are impossible to measure
under usual circumstances.
Thus, using formulas in \cite{my-prd-07}, we find that PV parts of rates are proportional to
\begin{eqnarray}
&&
\int d{\cal P}_{\nu} \sum_{h_k} \Re (A_0 \vec{A}^*) \propto
\int d{\cal P}_{\nu} \sum_{h_k} ( \frac{\vec{p}_i}{ E_i} + \frac{\vec{p}_j}{ E_j}) = 
\vec{k}  \frac{J_{ij}(\omega)}{\omega}
 \,,
\\ &&
 d{\cal P}_{\nu} = \frac{d^3p_i d^3p_j}{(2\pi)^2}\delta (\omega + E_i + E_j - \epsilon_{eg}) 
\delta (\vec{k} + \vec{p}_i + \vec{p}_j) 
\,.
\end{eqnarray}
The photon momentum vector $\vec{k}$ is thus
multiplied to give PV operator of the form, $\vec{k}\cdot\vec{\sigma}$
where $\vec{\sigma}$ is the electron spin operator $\times 2$.
The explicit form of function $J_{ij}(\omega)$ is given in 
eq.(\ref{vector integral}) of Appendix A.

This conclusion is consistent with
the ordinary view that PV effects must arise from interference
of parity odd combination of $V\cdot A$ in the product of electron
and quark 4-currents.
The spin current of electron $\propto \vec{\sigma}$ arises from
the spatial component of 4-axial vector $A \propto \gamma^{\alpha}\gamma_5$
in the non-relativistic limit,
while the nuclear mono-pole current $\propto j_q^0$
arises from the time component of 4-vector current $V \propto \gamma_{\alpha}$.
It is the unique combination of electron and nuclear current
operators 
that gives rise to large PV terms without the suppression of
$1/$mass order, which became possible only with the advent
of nuclear mono-pole contribution given in \cite{ys-13}.

Alkaline earth atoms are two-electron system of the 
angular momentum combination of parity odd orbitals, $sp$.
This combination of angular momenta
appears as the first excited group of levels
in alkaline earth atoms.
Two electrons may be either in the spin triplet or the spin
singlet state in the terminology of the $LS$ coupling scheme.
Thus, one has four different states (with the usual magnetic
degeneracy of energies), $^3\!P_2, ^3\!P_1, ^3\!P_0, ^1\!P_1$,
the atomic notation of $^{2S+1}L_J$ being used \cite{atomic physics}.

Another important consideration is that it is better
to use heavy (large atomic number) atoms for
large RENP rates \cite{ys-13}.
This poses a problem of state mixing in the $LS$ scheme,
which requires the use of intermediate coupling scheme
\cite{condon-shortley}.
The $LS$ coupling scheme is based on
the assumption that electrostatic interaction between electrons is
much larger than the spin-orbit interaction
$\sum_i \xi(r_i) \vec{l}_i\cdot\vec{s}_i$, which
however becomes larger for heavier atoms.
In the heaviest atoms such as Pb the $jj$ coupling scheme
becomes a better description \cite{condon-shortley}, 
but most of heavy atoms is well described
by the intermediate coupling scheme using the $LS$ basis.

In the intermediate coupling scheme applied to heavy
alkaline earth atoms one considers the mixing among states of
the same total angular momentum, since the total 
angular momentum is conserved under the presence of
the spin-orbit interaction.
These are $^3P_1$ and $^1P_1$ in the $LS $ scheme.
Energy eigenstates are given in terms of the $LS$ basis
\cite{standard notation},
\begin{eqnarray}
&&
|{}^+P_1 \rangle = \cos \theta |^1P_1 \rangle + \sin \theta |{}^3P_1 \rangle
\,, \hspace{0.5cm}
|{}^-P_1 \rangle = \cos \theta |{}^3P_1 \rangle - \sin \theta |{}^1P_1 \rangle
\,,
\end{eqnarray}
(with $\pm$ denoting larger/smaller energy state)
where the angle $\theta$ is determined by
the strength of spin-orbit interaction in the system
and is related to experimental data
of level energies.
In the Yb case $\sin \theta \sim 0.16$ \cite{ysu-pv-13}.
Dipole moments $d(|{}^{\pm}P_1 \rangle \rightarrow |{}^1S_0\rangle$ 
needed for RENP calculation
are induced by a non-vanishing value of $\theta$.

\begin{figure*}[htbp]
 \begin{center}
 \epsfxsize=0.6\textwidth
 \centerline{\epsfbox{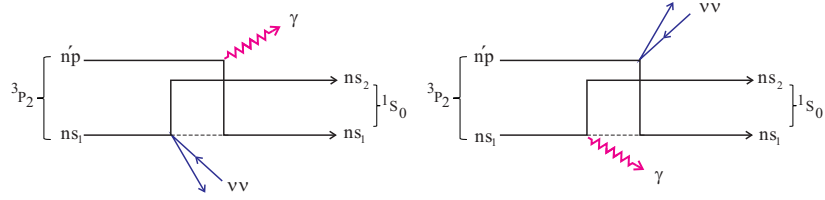}} \hspace*{\fill}
   \caption{Parity odd contribution of valence electron exchange.
   Neutrino pair emission contains the
   PE part of vertex, as described in the text.
 }
   \label {core renp pcpv 2}
 \end{center} 
\end{figure*}

We now turn to a concrete explanation of how PO amplitude arises,
corresponding to the left diagram of Fig(\ref{core renp pcpv 2}).
An electron in the $ns_1$ level of the two-electron system of
excited $ns_1, n'p$ state first makes a virtual transition 
to a vacant level in $ns_2$ 
by neutrino pair emission operator 
$\propto e^{\dagger}\vec{\sigma}e \cdot
\nu_i^{\dagger} \vec{\sigma}\nu_j$.
Another electron in the excited level $n'p$ then 
fills the hole in $ns_1 $ by a photon emission,
completing the transition 
$|ns_1 n'p \rangle \rightarrow | ns_2 ns_1 \rangle 
+ \gamma+\nu_i \nu_j$. 
One might think that another, equally contributing 
possibility is a process in which the neutrino pair emission
and the photon emission vertexes are interchanged in the time sequence.
This is the diagram in the right of Fig(\ref{core renp pcpv 2}),
but the quantum numbers of two-electron system
changes according to ${}^3P_2 \rightarrow {}^{\pm}P_1 \rightarrow {}^1S_0$,
thus this contribution is highly forbidden both by
E1 and the spin operators involved.

\begin{figure}[htbp]
\begin{minipage}{8cm}
	\epsfxsize=\textwidth
	\centerline{\epsfbox{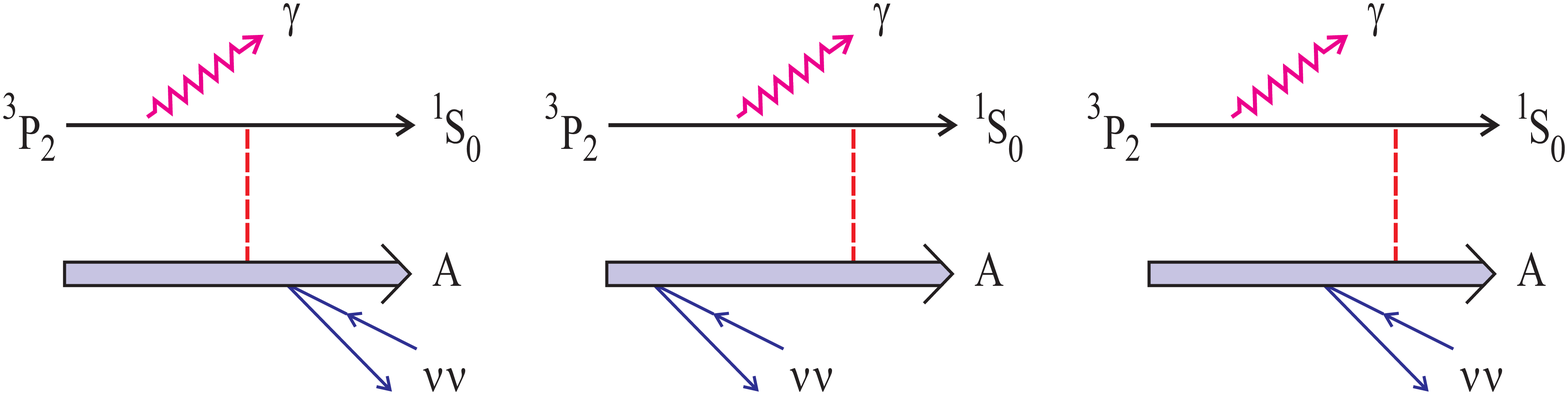}} \hspace*{\fill}
   \caption{Rejected PE diagrams that give vanishing contribution.
}
   \label{pc diagrams without b 1}
\end{minipage}
\begin{minipage}{1.5cm} $\;$ \end{minipage}
\begin{minipage}{8cm}
	\epsfxsize=\textwidth
	\centerline{\epsfbox{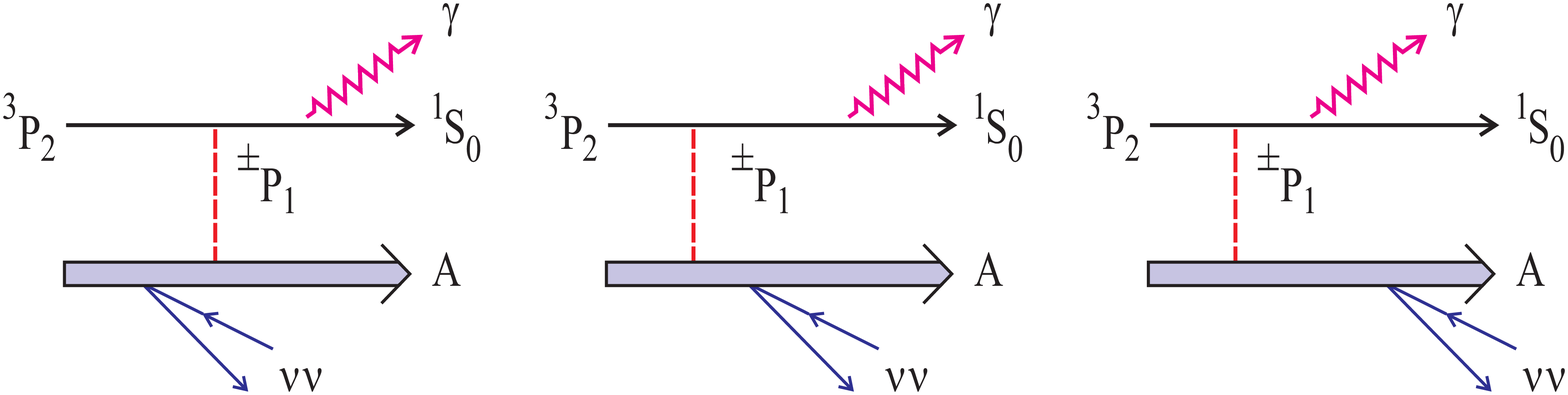}} \hspace*{\fill}
   \caption{Candidate PE diagrams.
}
   \label{pc diagrams without b 2}
\end{minipage}
\end{figure}

We now turn to PE amplitude that may have a large interference with
this PO amplitude.
In a recent work \cite{ys-13} we discussed a possibility
of  largest PC rate using the nuclear mono-pole current
(time component of 4-vector part) for neutrino pair emission.
A candidate set of PE amplitude might arise from diagrams of
Fig(\ref{pc diagrams without b 1}) and 
Fig(\ref{pc diagrams without b 2}).
The neutrino pair emission $\propto Q_w$ occurs
from the nuclear line, and the rest consists of
the Coulomb interaction $\propto Z\alpha/r$ and E1 emission.
The quantum numbers of atomic transition,
${}^3P_2 \rightarrow {}^{\pm}P_1 \rightarrow {}^1S_0$,
dictates the time sequence of the Coulomb interaction first
and E1 emission next, 
thus rejecting the possibility of Fig(\ref{pc diagrams without b 1}).

Contribution from Fig(\ref{pc diagrams without b 2})
is calculated as follows.
Combined with the time of nuclear pair emission,
there are three types of diagrams giving different energy denominators.
Each of these contain numerator factors of the form,
\begin{eqnarray}
&&
\langle {}^1S_0| \vec{E}\cdot \vec{D} | {}^{\pm}P_1 \rangle
\langle {}^{\pm}P_1 | \frac{Z\alpha}{r} | {}^3P_2 \rangle
= \pm \sin \theta \cos \theta 
\langle {}^1S_0| \vec{E}\cdot \vec{D} | {}^1P_1 \rangle
\langle {}^3P_1 | \frac{Z\alpha}{r} | {}^3P_2 \rangle
\equiv \pm {\cal N}_0
\,.
\end{eqnarray}
Amplitudes consist of six terms, considering different
$|^{\pm}P_1 \rangle$ intermediate states.
Three contributions from each of $|{}^{\pm}P_1 \rangle$,
using the energy conservation $\epsilon({}^3P_2)= E_{2\nu} + \omega$, 
add to a common factor $\pm {\cal N}_0$ times
\begin{eqnarray}
&&
\frac{1}{(\epsilon_3 - \omega)(\epsilon_{\pm}-\omega)} 
- \frac{1}{(\epsilon_{\pm}- \epsilon_3 \,)(\epsilon_3 - \omega)} 
+ \frac{1}{(\epsilon_{\pm}- \epsilon_3\,)(\epsilon_{\pm}-\omega)}
= 0
\,, 
\end{eqnarray}
with $\epsilon_3 = \epsilon({}^3P_2)- \epsilon({}^1S_1)$.
Thus, we conclude that the lowest order contribution 
given by Fig(\ref{pc diagrams without b 1})
and Fig(\ref{pc diagrams without b 2}) to PE amplitude
vanishes and the magnetic field assistance as
described in the next section is required for non-vanishing contribution.

We shall not apply external static electric
field, because it may induce an instrumental
parity mixture difficult to disentangle from the intrinsic
parity violation of fundamental theory \cite{yfsy-10}.

\vspace{0.5cm}
\lromn4 
{\bf Zeeman mixing and magnetic factors}

\vspace{0.3cm}

The Zeeman mixing caused by the magnetic field 
is described by the interaction vertex 
$\mu_B (2\vec{S}+\vec{L})\cdot \vec{B}$ \cite{atomic physics}.
This Zeeman mixing applied to our problem gives
perturbed states,
\begin{eqnarray}
&&
| e \rangle' = |e \rangle + \delta_{+e} | ^{+}{}P_1 \rangle
+ \delta_{-e} | ^{-}{}P_1 \rangle
\,, \hspace{0.5cm}
\delta_{\pm e} = 
\frac{\langle ^{\pm}{}P_1 |\mu_B(2\vec{S} + \vec{L})\cdot 
\vec{B}|e\rangle}{\epsilon_{\pm e}}
\,.
\end{eqnarray}
The mixing amplitude $\delta_{\pm e}$, 
with $\mu_B B \sim 5.8 \times 10^{-5} $eV/T,  
gives a small, but important transition between
different $J$ states.
With the Zeeman mixing inserted in diagrams of Fig(\ref{core renp pcpv 1}),
the product of atomic matrix elements ${\cal N}_0$ above is modified to 
\begin{eqnarray}
&&
\pm {\cal N} \,, \hspace{0.5cm}
{\cal N} =
\sin \theta \cos \theta 
\langle {}^1S_0| \vec{E}\cdot \vec{D} | {}^1P_1 \rangle
\langle {}^3P_1 | \mu_B (2\vec{S} + \vec{L})
\cdot\vec{S}| {}^3P_2 \rangle
\langle nP| \frac{Z\alpha}{r}|nP \rangle
\,. 
\end{eqnarray}
The last factor $\propto Z\alpha/r$ of Coulomb energy is estimated using
Thomas-Fermi model as done in \cite{ys-13}, giving
$\sim 31 {\rm eV} Z^{4/3}$.

\begin{figure*}[htbp]
 \begin{center}
 \epsfxsize=0.6\textwidth
 \centerline{\epsfbox{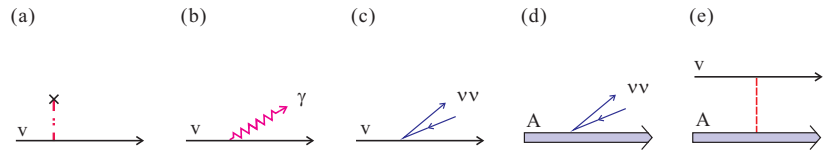}} \hspace*{\fill}
   \caption{Five basic units of interaction.
Cross is for Zeeman field, dotted line for instantaneous Coulomb interaction.
v means the valence electron line and A is atomic nucleus.
}
   \label {core renp elements}
 \end{center} 
\end{figure*}

\begin{figure*}[htbp]
 \begin{center}
 \epsfxsize=0.6\textwidth
 \centerline{\epsfbox{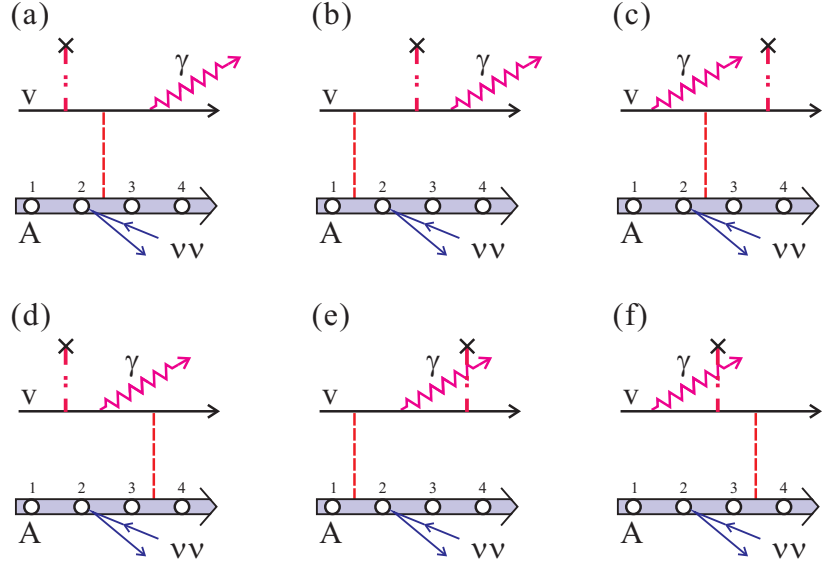}} \hspace*{\fill}
   \caption{24 PC RENP diagrams.
   Along the nuclear line neutrino pair emission may occur
   in four places in time sequence relative to three vertexes 
   along the valence line,
   four different nuclear vertex locations giving different amplitudes.
In our 3-level approximation only (a) and (c) contribute.
 }
   \label {core renp pcpv 1}
 \end{center} 
\end{figure*}

We now turn to detailed description of this unique 
candidate for PV effect.
There are five vertexes to be considered and
we shall treat these basic interaction units as shown in
Fig (\ref{core renp elements}) on an equal footing.
Five types of interactions have to be considered;
valence transition by Zeeman field  $\mu_B (2S+L) \cdot B$ 
of Fig (\ref{core renp elements}) (a),
E1 photon emission $d\cdot E$ (b), 
neutrino pair emission from valence electron which leads to
parity violation (c),
neutrino pair emission from nucleus (by the mono-pole current
as described in \cite{ys-13})  (d), 
 and Coulomb interaction between valence
electron and nucleus $V_C$ (e).

Five units of interaction along
the valence electron line are given by five vertex matrix elements
of operators,
\begin{eqnarray}
&&
\hspace*{-1cm}
\mu_B  (2\vec{S} + \vec{L})\cdot\vec{B} \,, \hspace{0.3cm}
\vec{d}\cdot\vec{E}
\,, \hspace{0.3cm}
a_{ij} \vec{\sigma}_e \nu_i^{\dagger} \vec{\sigma} \nu_j
\,, \hspace{0.3cm}
Q_w\nu_i^{\dagger}\nu_j \,,\;
(Q_w \equiv  N - 0.044 Z)
\,, \hspace{0.3cm}
\frac{Z \alpha}{r}
\,.
\end{eqnarray}
RENP amplitudes consist of factors of these basic units,
energy denominators in perturbation theory, and
coupling factors.
Neutrino pair emission gives rise to product of two plane wave
functions of definite helicities.
For Majorana pair emission the wave function of
two neutrinos must be anti-symmetrized, since
Marjorana particles are identical to
their own anti-particles and effects of identical
fermions work, to give rise to the principle
of Majorana-Dirac distinction \cite{my-prd-07}.

It is important, and experimentally useful, 
to work out effects of magnetic field directional dependence.
This magnetic field dependence of amplitudes and rates is called
the magnetic factor generically in the following.
We consider the experimental setup in which
a static magnetic field is applied in a general direction
tilted by an angle $\theta_m$ from the trigger z-axis
(which is also the direction of emitted photon).
Magnetic quantum numbers $M$ of states are defined as
components of $\vec{J}$ along the quantization axis,
namely the magnetic field direction.
To emphasize directionality we denote states by
the notation of tilde, hence
\begin{eqnarray}
&&
\widetilde{ |J, M \rangle } 
= e^{-i\theta_m J_y} | J, M \rangle
= \sum_{M'} d^J_{M,M'}(\theta_m) | J, M' \rangle
\,,
\end{eqnarray}
where $d^J_{M,M'}(\theta_m)$ is the Wigner d-function 
or the rotation matrix in the terminology of \cite{rose}.

Let us first work out the magnetic factor associated with
the PE (parity even) amplitude.
The magnetic factor for emission of the photon
circular polarization $h=\pm$ is given by
\begin{eqnarray}
&&
\sum_M \langle {}^1S_0 | er Y_{1, \pm 1}
\widetilde{ | {}^{\pm}P_1 M \rangle}
\langle  {}^{\pm}P_1\tilde{M} 
|(2\tilde{ S}+\tilde{L} )_z
\widetilde{ | {}^3P_2 M' \rangle}
\,,
\\ &&
\sum_M \langle {}^1S_0 | er Y_{1, \pm 1}
\widetilde{ | {}^{\pm}P_1 M \rangle} 
\widetilde{\langle  {}^{\pm}P_1 M |} (2\tilde{ S}+\tilde{L})_z
\widetilde{| {}^3P_0 M' \rangle}
\,.
\end{eqnarray}
The operator $er Y_{1, \pm}$ is the atomic dipole transition
operator for emission of the specified photon circular polarization $h=\pm$.
Since summation over magnetic quantum numbers in intermediate
states can be taken along any axis, we took the axis
along the magnetic field, which makes calculations easier.
(The magnetic quantum number in the initial state is taken
along with the magnetic field, which is dictated in the
experimental setup.)

The magnetic field mixes states of ${}^3P_{2,0}$ and ${}^{\pm}P_1$
by the atomic operator $2\vec{S} + \vec{L} = \vec{J} + \vec{S}$.
The total angular momentum $\vec{J}$ here does not contribute
since $\Delta J \neq 0$ in two involved states.
This implies that only ${}^3P_1$ components of  ${}^{\pm}P_1$
have non-vanishing matrix element of
\begin{eqnarray}
&&
\widetilde{\langle {}^3P_1,  M_J |} 
(2\tilde{\vec{S}} + \tilde{\vec{L}})_q
\widetilde{ | {}^3P_2 \rangle} 
\langle {}^3P_1,  M_J | (2\vec{S} + \vec{L})_q | {}^3P_2 \rangle
= \sqrt{\frac{5}{2}}  (-1)^{1-M_J} 
\left(
\begin{array}{ccc}
1 & 1  & 2 \\
-M_J  &  q & M_J-q
\end{array}
\right)
\end{eqnarray}
A similar relation exists for the transition from
$^3P_0$.
Reduced matrix element,
$\langle {}^3P_1 || \vec{S} || {}^3P_2 \rangle = \sqrt{5/2}$ was used.
Thus, the magnetic factor associated with PE amplitude is given by
\begin{eqnarray}
&&
\sum_M \langle {}^1S_0 | Y_{1, \pm 1}
\widetilde{ | {}^{\pm}P_1 M \rangle}
\langle  {}^{\pm}P_1\tilde{M} 
|(2\tilde{ S}+\tilde{L} )_z
\widetilde{ | {}^3P_2 M' \rangle}
\nonumber \\ &&
= \pm \sqrt{\frac{5}{2}} \langle {}^1S_0|er | {}^1P_1 \rangle \sin \theta \cos \theta 
d^1_{M', \mp 1}(-1)^{1-M'} 
\left(
\begin{array}{ccc}
1 & 1  & 2 \\
-M'  &  0 & M'
\end{array}
\right)
\,,
\\ &&
\sum_M \langle {}^1S_0 | Y_{1, \pm 1}
\widetilde{ | {}^{\pm}P_1 M \rangle} 
\widetilde{\langle  {}^{\pm}P_1 M |} (2\tilde{ S}+\tilde{L})_z
\widetilde{| {}^3P_0 M' \rangle}
\nonumber \\ &&
= \pm  \sqrt{\frac{5}{2}}  \langle {}^1S_0|er | {}^1P_1 \rangle 
\sin \theta \cos \theta d^1_{M', \mp 1}(-1)^{1-M'} 
\delta_{M', 0}
\left(
\begin{array}{ccc}
1 & 1  & 0 \\
0  &  0 & 0
\end{array}
\right)
\,.
\end{eqnarray}
We may define the magnetic factors for amplitudes
by extracting out dipole matrix element 
$\langle {}^1S_0|er | {}^1P_1 \rangle 
\sin \theta \cos \theta$, which is related to
measured A-coefficient and energy difference of atomic levels,
The magnetic factor for ${}^3P_2$ is
\begin{eqnarray}
&&
W_{1, \pm}^M(x) = \sqrt{\frac{5}{2}} (-1)^{1-M} 
\left(
\begin{array}{ccc}
1 & 1  & 2 \\
-M  &  0 & M
\end{array}
\right)
d^1_{M, \mp 1}(x)
\,.
\end{eqnarray}

Similar magnetic factor for PO amplitude is
defined by taking into account of the neutrino
phase space integration which gives
$\vec{k}$, the wave vector of emitted photon. 
It is for $^3P_2$ RENP
\begin{eqnarray}
&&
\sum_M \langle {}^1S_0 | Y_{1, \pm 1}
\widetilde{ | {}^{\pm}P_1 M \rangle}
\widetilde{\langle  {}^{\pm}P_1 M}
|(2S+L )_z
\widetilde{| {}^3P_2 M' \rangle}
\,.
\end{eqnarray}
Note that the definite field direction along the
trigger axis (fixed as parallel to $z$ axis) is selected,
hence no tilde operation in this formula of angular momenta.
Thus, the magnetic factor for PO is more complicated;
\begin{eqnarray}
&&
W_{2, \pm}^M(x) = - \sqrt{\frac{5}{2}}
\sum_{M_1, M_2} 
(-1)^{1-M_1} 
\left(
\begin{array}{ccc}
1 & 1  & 2 \\
-M_1  &  0 & M_1
\end{array}
\right)
d^2_{M, M_1}(x) d^1_{M_2, M_1}(x) d^1_{M_2, \pm 1}(x)
\,.
\end{eqnarray}
Explicit forms of these functions are given in Appendix B.
They are simple linear combinations of sinusoidal functions.

PV odd rates are given by differences of
the product of magnetic factors for PO and PE amplitudes.
It turns out that the PO product magnetic factor for
$^3P_0$ RENP vanishes, and we shall work out
quantities for $^3P_2$ RENP in the following.
There are two kinds of PV asymmetries one can
calculate from these magnetic factors:
the first one is PV asymmetry under the magnetic field
reversal, $x\rightarrow \pi -x$,
and the other is the asymmetry under
the reversal of the photon
circular polarization, $h=\pm \rightarrow \mp$,
for which all angle dependences may be integrated out.
PV asymmetry under field reversal is dictated by
the magnetic factor,
\begin{eqnarray}
&&
{\cal M}^M(x) \equiv \sum_{\pm} {\cal M}^M_{\pm}(x)
\,, \hspace{0.5cm}
{\cal M}^M_{\pm}(x) =
W_{1, \pm }^M(x)W_{2, \pm}^M(x)  
- W_{1, \pm}^M(\pi - x)W_{2, \pm}^M(\pi - x)
\,.
\end{eqnarray}
Explicitly worked out, these are
\begin{eqnarray}
&&
{\cal M}^{\pm 1}(x) = - \frac{1}{2}\cos^3 x
\,, \hspace{0.5cm}
{\cal M}^{0}(x) =  \sin^2 x \cos x
\,.
\end{eqnarray}
Non-vanishing values at various angles
may be taken as indication of parity violation in RENP.
The simplest PV asymmetry of this kind is the forward-backward asymmetry
given by ${\cal M}_1^{0}(0) = 0$ and ${\cal M}_1^{\pm}(0) = - \frac{1}{2}$.
For normalized asymmetries rate differences should
be divided by PE combinations of angular factors,
\begin{eqnarray}
&&
{\cal M}_1^{M}(x) \equiv 
\sum_{\pm} (W_{1, \pm }^M(x)\,)^2   
+ (W_{1, \pm}^M(\pi - x)\,)^2 
\,,
\\ &&
{\cal M}_2^{M}(x) \equiv
\sum_{\pm} W_{1, \pm }^M(x)W_{2, \pm}^M(x)  
+ W_{1, \pm}^M(\pi - x)W_{2, \pm}^M(\pi - x)
\,.
\end{eqnarray}
Explicit forms of these are listed in Appendix B.

The other PV asymmetry under the reversal of the photon circular
polarization is given by
\begin{eqnarray}
&&
\int_{-\pi}^{\pi} dx \left( {\cal M}_+^{\pm 1} (x) - {\cal M}_-^{\pm 1} (x)
\right) = \pm 0.39 \,, \hspace{0.5cm}
\int_{-\pi}^{\pi} dx \left( {\cal M}_+^{0} (x) - {\cal M}_-^{0} (x)
\right) = 0
\,.
\label{pv asym under cir pol}
\end{eqnarray}

\begin{figure*}[htbp]
 \begin{center}
 \epsfxsize=0.6\textwidth
 \centerline{\epsfbox{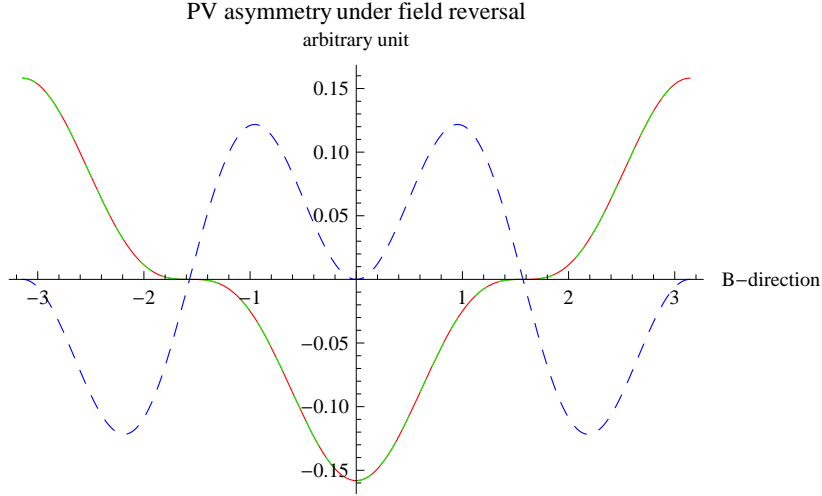}} \hspace*{\fill}
   \caption{$^3P_2 \rightarrow ^1\!S_0$ PV asymmetry under field reversal
for the sum of two circular polarizations   
vs B-direction measured from the trigger axis.
Initial magnetic quantum number of $M=\pm 1 $ (the degenerate case)
 is depicted in solid red and dash-dotted green,
and $M=0$ in dashed blue.
}
   \label{magnetic factor 2->0}
 \end{center} 
\end{figure*}

These magnetic factors are 
plotted for magnetic quantum numbers of $M= \pm 1, 0$ in 
Fig(\ref{magnetic factor 2->0}).
Directional dependence of PV asymmetries is large and should
help much in proving the weak origin of RENP process.

\vspace{0.5cm}
\lromn5
{\bf PV interference, PC rate and PV asymmetry}

\vspace{0.3cm}
RENP spectral rates may be expressed by two
formulas $\Gamma_{2\nu\gamma}^{\pm}(\omega)$ 
which are interchanged by reversal of instrumental
polarity; the magnetic field direction and the direction
of circular polarizations.
Rates may be written as
\begin{eqnarray}
&&
\Gamma_{2\nu\gamma}^{\pm}(\omega) = \Gamma_{2\nu\gamma}^{PC1}(\omega)
+ \Gamma_{2\nu\gamma}^{PC2}(\omega) \pm \Gamma_{2\nu\gamma}^{PV}(\omega)
\,.
\end{eqnarray}
The last term is the interference term arising from
the product of PE and PO amplitudes, while
the first two terms result from the squared PE and PO amplitudes.
We decompose these three spectral rates, 
both parity conserving (PC) and parity violating (PV),
into an overall factor denoted by $\Gamma_0$, 
various spectral shape functions of kinematical nature,
atomic factors, and the dynamical factor $\eta_{\omega}(t)$.
We shall use a unit of 100 MHz for A-coefficients (decay rates)
and eV for all energies.
We give rates appropriate for Yb $J=2 \rightarrow 0$ RENP.
The conversion factor in our natural unit
is $\hbar c = 1.97 \times 10^{-5} {\rm eV}\cdot{\rm cm}$.

The overall rate is given by
\begin{eqnarray}
&&
\Gamma_0 = 
\frac{3}{4} G_F^2 \epsilon_{eg} n^3 V  
\frac{\gamma_{+g}}{\epsilon_{+g}^3} (\sin \theta \cos \theta)^2
\eta_{\omega}(t)
\\ &&
\sim 54 {\rm mHz} (\frac{n}{10^{21}{\rm cm}^{-3}})^3 \frac{V}{10^2 {\rm cm}^3} 
\frac{\epsilon_{eg}}{{\rm eV}} 
\frac{\gamma_{pg}{\rm eV}^{3}}{\epsilon_{pg}^3{\rm 100 MHz}} 
(\sin \theta \cos \theta)^2 \eta_{\omega}(t)
\,.
\label{overall rate}
\end{eqnarray}
The factor $\sin \theta \cos \theta$ reflects 
the strength of the spin-orbit interaction in heavy atoms.
As representative values of atomic data
we may take the dominant dipole strength 
$d_{pg} = \sqrt{3\pi \gamma_{pg}/\epsilon_{pg}^3}$,
of state $|p\rangle = ^+\!P_1$ for Yb.
Electric field strength of emitted photons 
has been written as $|E|^2 = \epsilon_{eg} n \eta_{\omega}(t)$
where $\epsilon_{eg} n$ is the maximum stored energy density stored
in the upper level $|e \rangle$.
Thus, one may regard $\eta_{\omega}(t)$ as the fraction of extractable energy
density within the target.
This quantity may be computed numerically using the PSR master equation 
\cite{ptep overview}.

Individual contributions are given as follows.
We present results for PV asymmetry under field reversal
using ${\cal M}_i = \sum_M {\cal M}_i^M (\theta_m)$ for the magnetic factor.
For the asymmetry under polarization reversal
this function should be replaced by the integrated 
quantity (\ref{pv asym under cir pol}).

(1) PC rate from squared PE amplitudes is given by
\begin{eqnarray}
&&
\Gamma_{2\nu \gamma}^{PC1} = \Gamma_0 
Q_w^2 V_C^2 
\left(\sum_{p=\pm}\epsilon_{pe} \delta_{pe}F_{C}(\omega; \epsilon_p)
\right)^2  
I(\omega)
{\cal M}_{1}(\theta_m) 
\,, \hspace{0.5cm}
I(\omega) =\sum_i I_{ii}(\omega)
\theta(\omega_{ii} - \omega)
\,,
\\ &&
I_{ii}(\omega) = 
\frac{1}{2}(\, C_{ii}(\omega)+ A_{ii}(\omega)
 + \delta_M m_1 m_2 D_{ii}(\omega) 
 \,)
\,, \hspace{0.5cm}
V_C \sim 31 {\rm eV} Z^{4/3}
\,, \hspace{0.5cm}
Q_w = N - 0.044 Z
\,,
\\ &&
F_{C}(\omega; \epsilon_p) =
\frac{1}{(\epsilon_{eg}-\omega )(\epsilon_{pg}-\omega)^2}
+ \frac{1}{\epsilon_{pe}(\epsilon_{pg}-\omega)^2}
+  \frac{1}{\epsilon_{pe}^2(\epsilon_{pg}-\omega)}
+  \frac{1}{\epsilon_{pe}^2(\epsilon_{pe}+\omega)}
\,.
\end{eqnarray}
We refer to Appendix A for all spectral shape functions here
and in the follwoing, \\
$ A_{ii}(\omega)\,, B_{ii}(\omega)\,,
C_{ii}(\omega)\,, D_{ii}(\omega)\,, J_{ii}(\omega)$ 
that arise from the neutrino phase space integration.

(2) 
PC rate arising from squared 
valence PO amplitude is 
\begin{eqnarray}
&&
\Gamma_{2\nu \gamma}^{PC2} = \Gamma_0 f_{vc}^2
H(\omega; \theta_m) {\cal M}_{2}(\theta_m)
\,, \hspace{0.5cm}
H(\omega; \theta_m) = \sum_i a_{ii}^2 H_{ii}(\omega)
\theta(\omega_{ii} - \omega)
\,,
\\ &&
H_{ii}(\omega) = \frac{1}{2} \left(
C_{ii}(\omega) - A_{ii}(\omega) - \delta_M m_i^2 D_{ii}(\omega)
\right) + \frac{B_{ii}(\omega)}{\omega^2}
\,, \hspace{0.5cm}
f_v(\omega) = - \frac{1}{\epsilon_{+g} -\omega}-
\frac{\gamma_{-g}\epsilon_{+g}^3}{\gamma_{+g}\epsilon_{-g}^3} 
\frac{1}{\epsilon_{-g} -\omega}
\,.
\end{eqnarray}

(3) Interference term between PO and PE amplitudes
is given by
\begin{eqnarray}
&&
\Gamma_{2\nu \gamma}^{PV} =  \Gamma_0 
Q_w f_{v}(\omega)V_C
\left(
\sum_{p=\pm}\epsilon_{pe}\delta_{pe}  F_C(\omega; \epsilon_p)
\right) J(\omega) 
{\cal M}(\theta_m)
\,,
\label{interference rate}
\\ &&
J(\omega) = \sum_i a_{ii} J_{ii}(\omega) \theta(\omega_{ii} - \omega)
\,,\hspace{0.5cm}
J_{ii}(\omega) = 
- \frac{\Delta_{ii}(\omega)}{4\pi}\omega
\left(
\epsilon_{eg} -\frac{4}{3} \omega
 +\frac{4 (\epsilon_{eg} - \omega) m_i^2 }{3\epsilon_{eg}  (\epsilon_{eg} -2 \omega) }
\right)
\,.
\end{eqnarray}
Note that three different magnetic factors,
${\cal M}\,, {\cal M}_{1,2}$, 
appear in three terms.

PV asymmetry is defined by
\begin{eqnarray}
&&
{\cal A}(\omega) = \frac{2 \Gamma_{2\nu \gamma}^{PV} }
{\Gamma_{2\nu \gamma}^{PC1} + \Gamma_{2\nu \gamma}^{PC2}}
\,.
\label{pv-asymmetry}
\end{eqnarray}
This is a quantity to be compared with
the experimental asymmetry obtained by
taking the ratio of the difference to the sum of two
rates when reversal of experimental setup
variables is made to change  instrumental parity.
The PV asymmetry ${\cal A}(\omega)$ of eq.(\ref{pv-asymmetry})
is a function of $M$ (the initial magnetic quantum number of
$^3P_2$ state) and $h$ the circular polarization.

\vspace{0.5cm}
\lromn6 
{\bf Numerical calculation of RENP spectral rates}

\vspace{0.3cm}
A-coefficients we need for computations of $^{174}_{70}$Yb RENP
are 
$\gamma_{+g} = 176, \gamma_{-g} =1.1$MHz's and
$\epsilon_{+g} =3.108, \epsilon_{-g} =2.2307, 
\epsilon(^3P_2) = 2.4438
$eV's.
The contribution of intermediates state
${}^+P_1$ dominates over ${}^-P_1$ with these parameters
due to larger values of $d^2 = 3\pi \gamma/\epsilon^3$;
$\gamma_{-g}\epsilon_{+g}^3/(\gamma_{+g}\epsilon_{-g}^3) 
\sim 0.017$
for Yb.
$\sin \theta \cos \theta \sim 0.158$ has been
estimated for Yb \cite{ysu-pv-13}.
The dominant Zeeman mixing 
is given by $\delta_{+e}$ with energy difference
$\epsilon_{+e} \sim 0.664$ eV.
Hence the magnetic mixing 
$\delta_{+e} = 5 \times 10^{-6}$ corresponds to
a magnetic field strength $\sim 57$ mT.
The nuclear electroweak is taken for even
isotope ${}^{174}$Yb, giving $Q_w \sim 101$.

\begin{figure}[htbp]
\begin{minipage}{8cm}
	\epsfxsize=\textwidth
	\centerline{\epsfbox{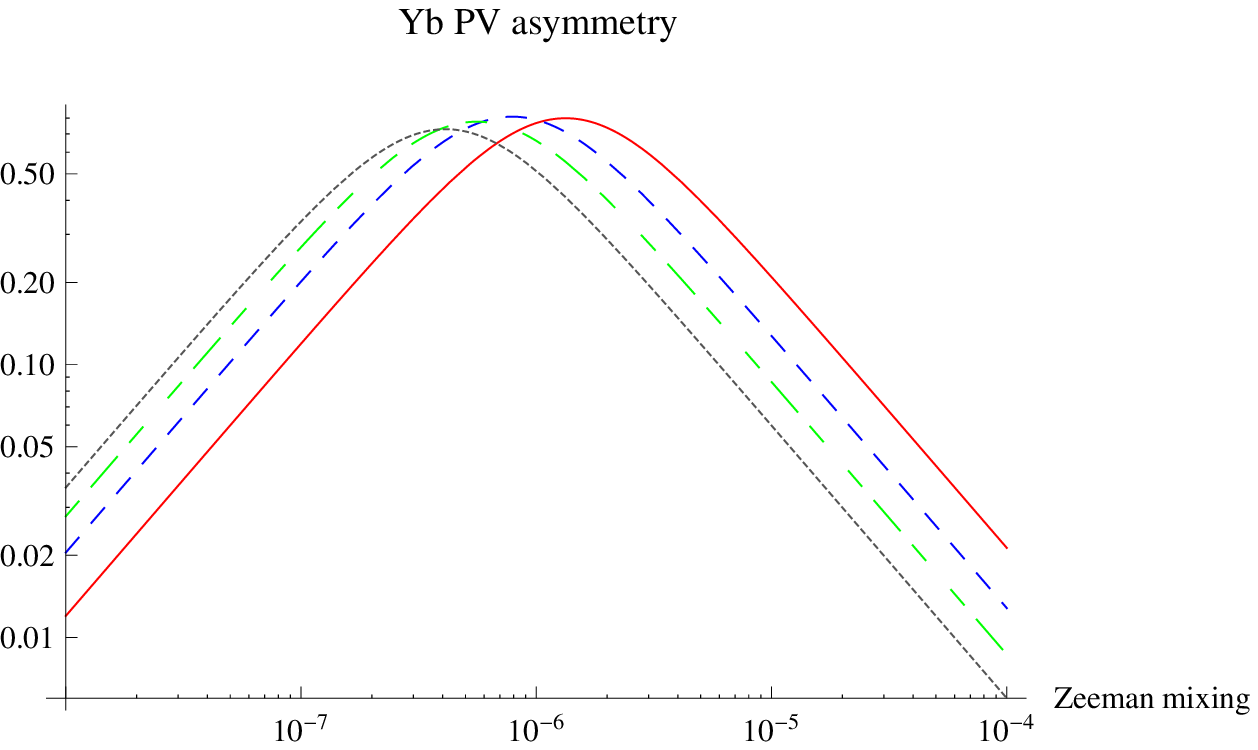}}
	\caption{Yb PV asymmetry under the 
reversal of photon circular
polarization plotted against the Zeeman
   mixing parameter $\delta_{+g}$, assuming a single neutrino
of mass 50 meV, the target number density $10^{22}$cm$^{-3}$,
and the target volume $10^2$cm$^3$.
Assumed photon energies are the level spacing of Yb 2.44 eV
$\times$ 0.1 in solid red, 0.2 in dashed blue, 0.3 in dash-dotted
green, and
0.4 in dotted black.}
	\label{pv asym vs mixing}
\end{minipage}
\begin{minipage}{2cm} $\;$ \end{minipage}
\begin{minipage}{8cm}
	\epsfxsize=\textwidth
	\centerline{\epsfbox{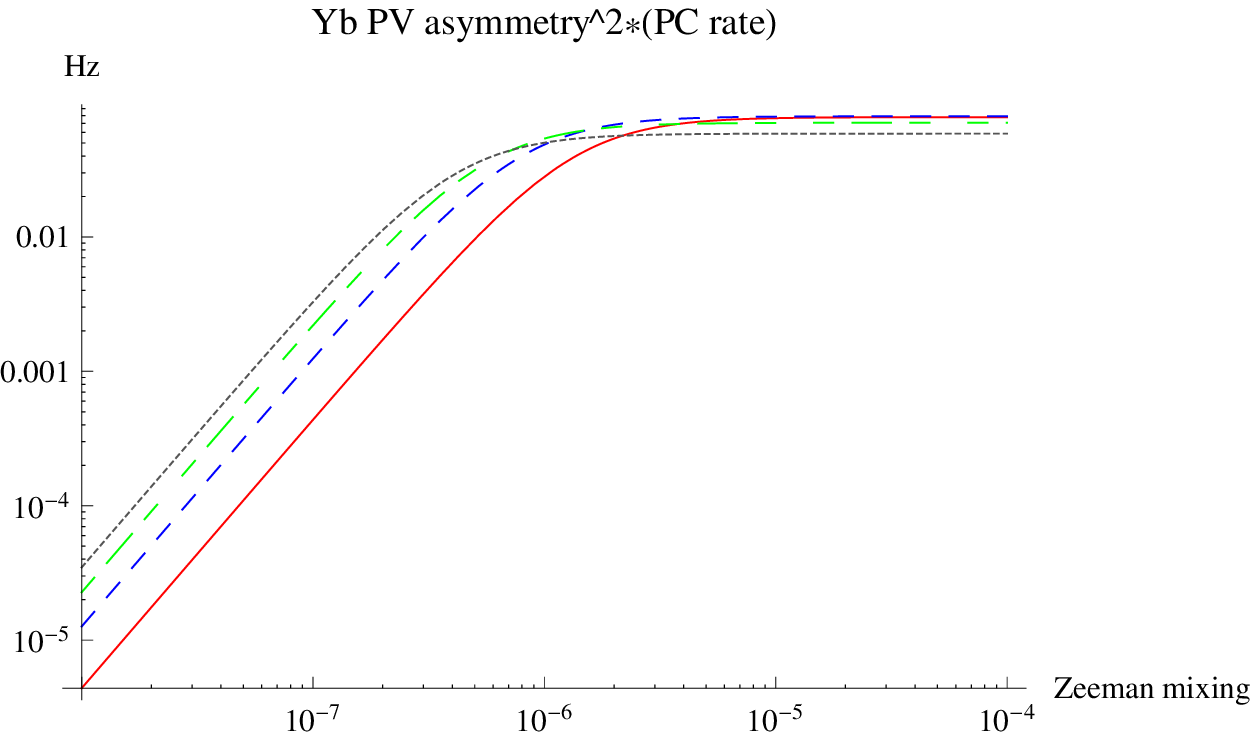}} \hspace*{\fill}
   \caption{Yb PV asymmetry squared $\times$
PC rate (figure of merits) plotted against the Zeeman
   mixing parameter $\delta_{+g}$, corresponding to
Fig(\ref{pv asym vs mixing}).
}
   \label{pv rate vs mixing}
\end{minipage}
\end{figure}

It is convenient to define a quantity which
may be called figure of merits;
the product of squared asymmetry times PC
rates.
This measures a statistical significance of
asymmetry measurements.
The figure of merits is plotted
against the magnetic mixing 
$\delta \sim 5 \times 10^{-5}$Tesla/eV,
in Fig(\ref{pv rate vs mixing}).
The magnitude of PV asymmetry under the reversal 
of circular polarization is shown in 
Fig(\ref{pv asym vs mixing}).
These results indicate that
there is an optimal choice of the magnetic
field strength, implying that a largest field
strength is not necessarily the best choice.
Based on this result we shall choose for the
following figures an optimal Zeeman mixing of $\sim 5 \times 10^{-6}$
which gives an optimal magnetic field strength $\sim 60$ mT.

\begin{figure*}[htbp]
 \begin{center}
 \epsfxsize=0.6\textwidth
 \centerline{\epsfbox{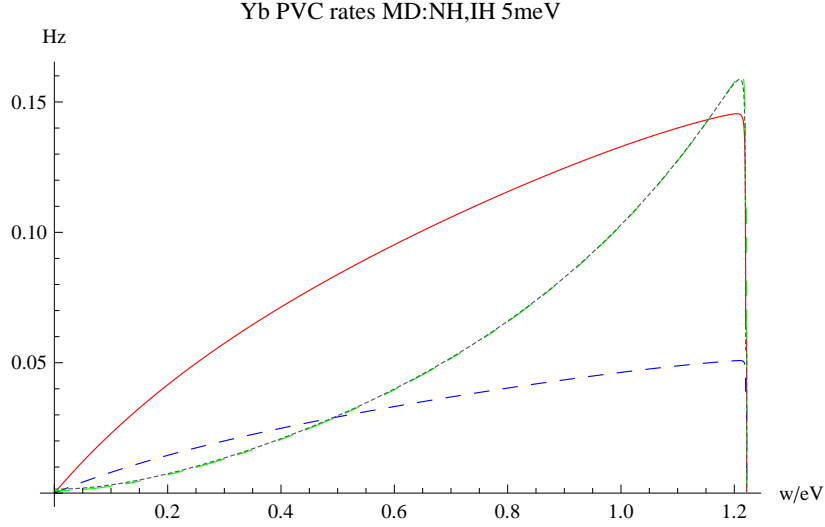}} \hspace*{\fill}
   \caption{${}^3P_2, J=2, M_J=1$ Yb PC rates, 
PV rate differences. Zeeman mixing amplitude $5\times 10^{-6}$
(corresponding to the magnetic field $\sim 60$ mT), $\eta_{\omega}(t) = 1$,
$n= 10^{22}$cm$^{-3}$, and $10^2$cm$^3$ are assumed.
Majorana NH PV in solid red, M-IH PV in dashed blue,
M-NH PC rate divided by 50 in dash-dotted green, 
and M-IH/50 in dotted
black (degenerate with M-NH PC). 
}
   \label{pvc rate 5}
 \end{center} 
\end{figure*}

\begin{figure*}[htbp]
 \begin{center}
 \epsfxsize=0.6\textwidth
 \centerline{\epsfbox{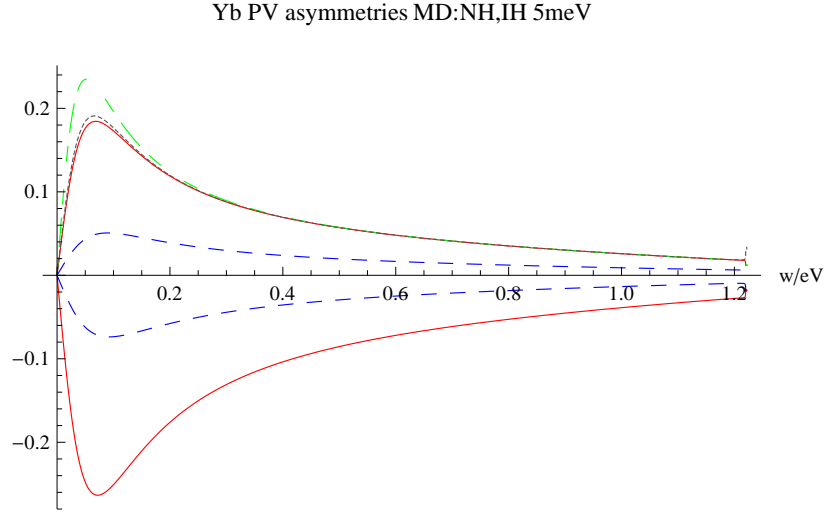}} \hspace*{\fill}
   \caption{${}^3P_2$Yb PV asymmetries vs photon energy.
Zeeman mixing amplitude $5\times 10^{-6}$, $\eta_{\omega}(t) = 1$,
$n= 10^{22}$cm$^{-3}$, and $10^2$cm$^3$ assumed.
In the positive side the Majorana case of
PV asymmetry under polarization reversal for NH 
is depicted in solid red,
M-IH case in dashed blue, D-NH in dash-dotted green and
the Dirac case for NH in dotted black.
In the negative side PV asymmetry under the field reversal
is plotted; M-NH in solid red, and M-IH in dashed blue,
all assuming the smallest
neutrino mass 5 meV.
}
   \label{pv asym 5}
 \end{center} 
\end{figure*}

\begin{figure*}[htbp]
 \begin{center}
 \epsfxsize=0.6\textwidth
 \centerline{\epsfbox{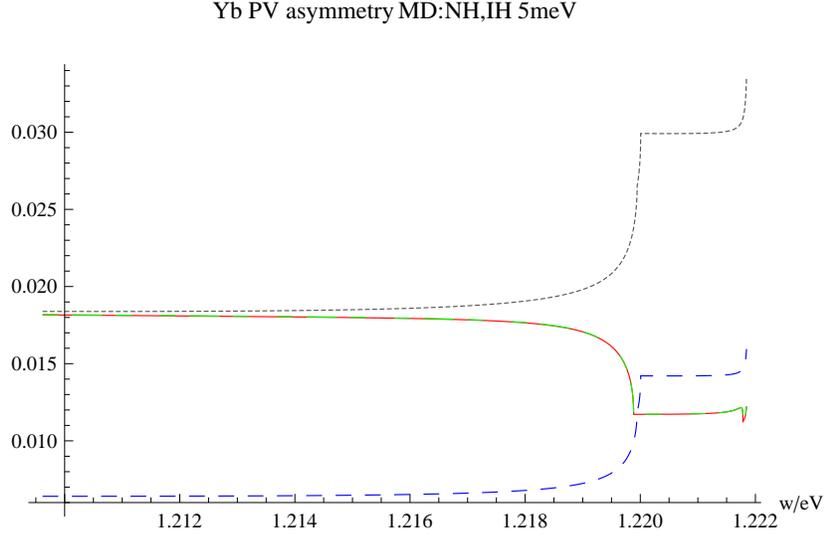}} \hspace*{\fill}
   \caption{${}^3P_2, J=2, M_J=1$ Yb PV asymmetry
 in the threshold regions
corresponding to Fig(\ref{pv asym 5}).
}
   \label{pv rate the 5}
 \end{center} 
\end{figure*}

In Fig(\ref{pvc rate 5}) $\sim$
 Fig(\ref{pv rate the 5}) 
we illustrate results of calculation
for RENP PV spectrum differences 
and PV asymmetry, assuming the smallest neutrino mass of
5 meV in which other neutrino parameters are
taken consistently with existing oscillation data.
In these and other figures a target number density
$n=10^{22}$cm$^{-3}$ and the target volume $V=10^2$cm$^3$
and the dynamical factor $\eta_{\omega}(t) = 1$
are taken, rates scaling with $n^3 V \eta_{\omega}(t)$.
Except in Fig(\ref{pv asym 5}) where
two different PV asymmetries are compared,
all other diagrams exhibit PV asymmetry
under the reversal of photon circular polarization.
Distinction of the normal hierarchical (NH) and the inverted
hierarchical (IH) mass patterns is easier for PV than
PC as seen in Fig(\ref{pvc rate 5}).
Overall PV rates for an optimal
magnetic field are typically of order $10^3$ larger than
hyperfine mixing in alkaline earth atoms of odd isotopes
given in \cite{ysu-pv-13}.

Dependence on the magnetic quantum number
$M$ of $J=2$ levels are as follows.
The magnitudes of PV asymmetries for $M= \pm 1$
are the same, while they vanish for $M=\pm 2, 0$.

Distinction of Majorana and Dirac neutrinos is of great interest.
Parity violating asymmetries do distinguish these two
cases when measurements by appropriate
choice of magnetic field $\approx $100 mT are made in the low photon
energies as evident in Fig(\ref{pv asym 5})
even for a smallest neutrino mass of 5 meV.

\begin{figure*}[htbp]
 \begin{center}
 \epsfxsize=0.6\textwidth
 \centerline{\epsfbox{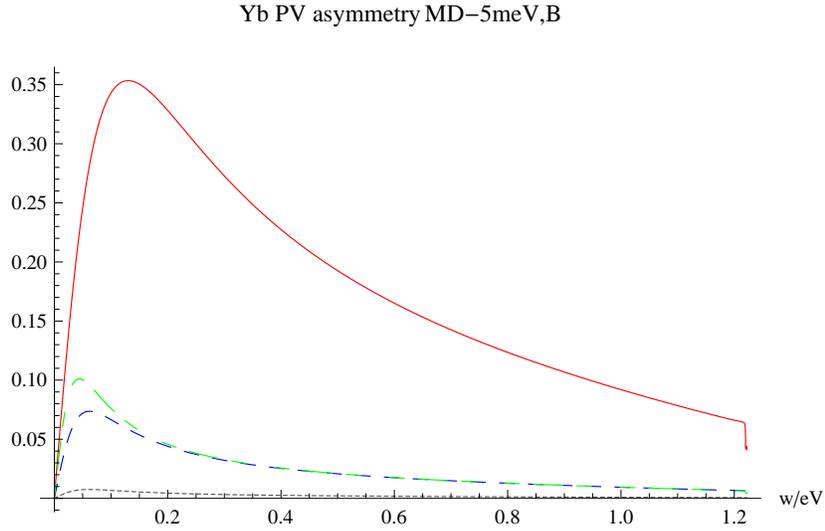}} \hspace*{\fill}
   \caption{${}^3P_2, J=2, M_J=1$ Yb 
PV asymmetries under the reversal of photon
circular polarization for
a few choices of magnetic fields, $B=10$mT
in solid red, $100$mT in dashed blue, $1$T in dotted black,
in the case of Majorana NH, 
and the Dirac NH case of $100$mT in dot-dashed green.
The assumed smallest neutrino mass is 5 meV.
}
   \label{pv rate bmag 5}
 \end{center} 
\end{figure*}

Fig(\ref{pv rate bmag 5}) shows dependence of PV asymmetry shapes
on the magnetic field strength for a few choices
of measured photon energies, which
clearly indicates the importance of
the field magnitude in actual experiments.

\begin{figure*}[htbp]
 \begin{center}
 \epsfxsize=0.6\textwidth
 \centerline{\epsfbox{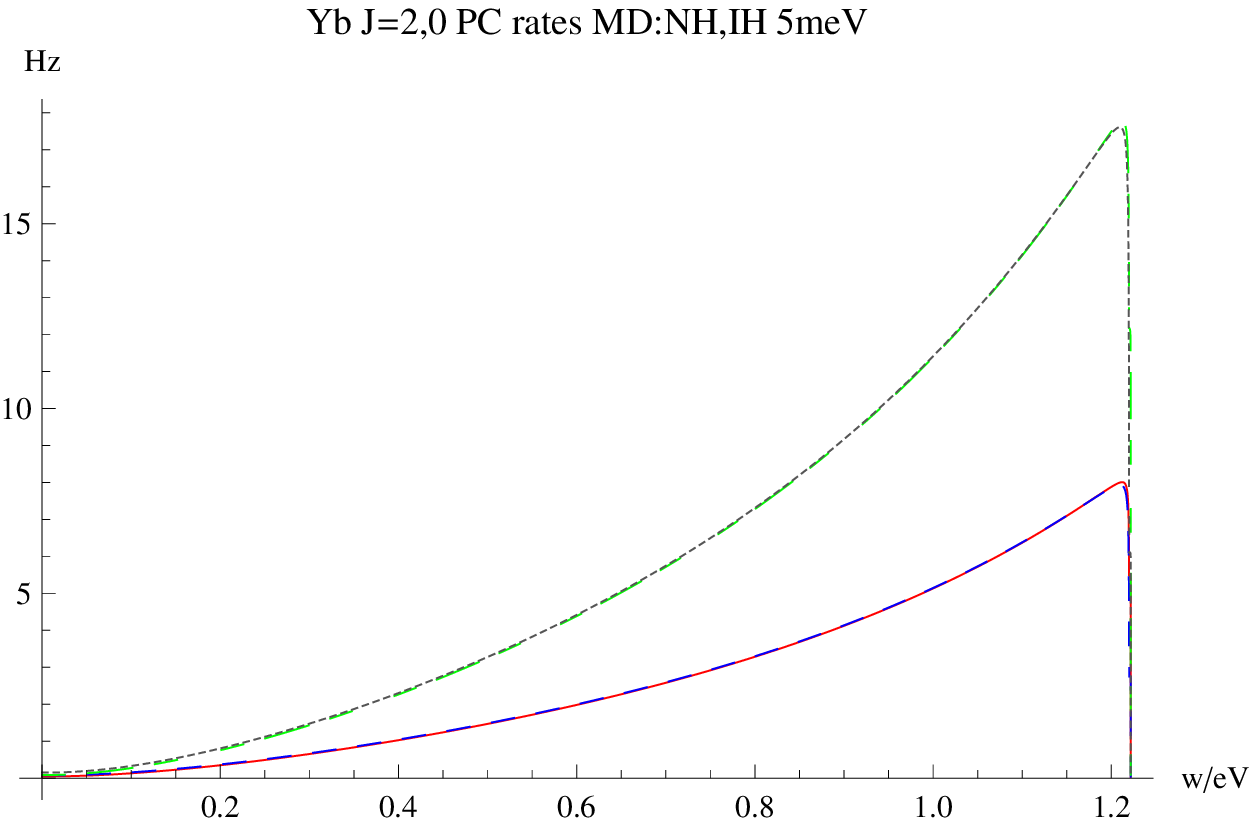}} \hspace*{\fill}
   \caption{Comparison of rates from ${}^3P_2, J=2, M_J=1$ 
and ${}^3P_0$ Yb PC rates, 
$\eta_{\omega}(t) = 1$,
$n= 10^{22}$cm$^{-3}$, and $10^2$cm$^3$ are assumed.
Majorana NH PC rate from ${}^3P_2$ in solid red, M-IH PC in dashed blue,
while Majorana NH PC rate from ${}^3P_0$ in dot-dashed green and
in dotted black.
}
   \label{pc20 rates 5}
 \end{center} 
\end{figure*}

Although parity violation effects do not exist for
${}^3P_0$ Yb RENP, it is of interest to
compare its PC rates with ${}^3P_2$ case.
This is shown in Fig(\ref{pc20 rates 5}).
In both cases NH and IH differences are small,
and difficult to resolve their differences in this figure.

Finally, we note that our method of computation
is readily applicable to other alkaline-earth-like atoms,
including an electron-hole system such as Xe excited
states of $6s 6p$ having the same quantum numbers
${}^3P_2$.

\vspace{0.5cm}
\lromn7
{\bf Summary}

\vspace{0.3cm}
We examined how parity violating
asymmetry and PV rate difference in RENP may be observed in atomic
de-excitation.
Our proposed mechanism uses interference terms
of parity even and odd amplitudes that do not
suffer from the usual atomic velocity suppression $v/c$,
since we use for the neutrino pair emission
the spin current contribution from the valence
electron and the nuclear mono-pole contribution
from nucleus.
Large PV interference and PV asymmetry
may occur in transitions among different parity states,
which suggests alkaline earth atoms as good targets.
Necessary state mixing between different $J$ states
occurs by an external magnetic field for alkaline
earth atoms of even isotopes.
Fundamental formulas applicable when
magnetic sub-levels are energetically resolved
are derived and used for numerical computations.
The PV asymmetry may readily reach of order several
tenths of unity
in the examined case of Yb.
Spectral shapes and PV asymmetries are sensitive
to the smallest neutrino mass, difference of the hierarchical
mass patterns, the Majorana-Dirac distinction.
Sensitivity to the applied magnetic field strength
may greatly help identification of RENP
process.
A further systematic search for better target atoms
of number density close to the Avogadro number per cm$^3$,
in particular  ions
implanted in transparent crystals,
is indispensable for realistic RENP experiments
along with extensive numerical simulations of
the time dependent dynamical factor ($\eta_{\omega}(t)$).

\vspace{0.5cm}
\lromn8 
{\bf Appendices}

\vspace{0.5cm}
{\bf Appendix A: Neutrino phase space integral}

\vspace{0.3cm}
Using the helicity summation formula of \cite{my-prd-07}
and disregarding irrelevant T-odd terms,
one has 
\begin{eqnarray}
&&
\hspace*{1cm}
\sum_{h_i} |j_0^{\nu}\cdot A_0 
+ \vec{j}^{\nu}\cdot\vec{A} |^2 = 
\nonumber \\ &&
\hspace*{-1cm}
\frac{1}{2} (1 + \frac{\vec{p}_1\cdot\vec{p}_2}{E_1E_2}+\delta_M 
\frac{m_1 m_2}{E_1E_2} ) 
|A_0|^2
+ 
\frac{1}{2} (1 - \frac{\vec{p}_1\cdot\vec{p}_2}{E_1E_2} -\delta_M 
\frac{m_1 m_2}{E_1E_2} ) 
|\vec{A}|^2 +  \frac{\Re(\vec{p}_1\cdot\vec{A} \vec{p}_2\cdot\vec{A}^*)}{E_1 E_2}
-2 (\frac{  \vec{p}_1}{ E_1} + \frac{  \vec{p}_2}{ E_2})  \Re (A_0 \vec{A}^*)
\,,
\nonumber \\ &&
\label{helicity summation of currents}
\end{eqnarray}
where $(E_i, \vec{p}_i)$ are neutrino 4-momenta.
In the phase space integral of neutrino momenta,
\begin{eqnarray}
&&
\int d{\cal P}_{\nu}(\cdots)
= 
\int \frac{d^3 p_1 d^3 p_2}{(2\pi)^2} \delta(E_1 + E_2 + \omega - \epsilon_{eg}) 
\delta(\vec{p}_1 + \vec{p}_2 + \vec{k}) (\cdots)
\end{eqnarray}
one of the momentum integration is used to eliminate the
delta function of the momentum conservation.
The resulting energy-conservation is used to fix the relative angle
factor $\cos \theta $
between the photon and the remaining neutrino momenta, 
$\vec{p}_1 \cdot \vec{k} = p_1 \omega \cos \theta$.
Noting the Jacobian factor $E_2/p\omega$
from the variable change to the cosine angle, one obtains one dimensional integral
over the neutrino energy $E_1$:
\begin{eqnarray}
&&
\frac{1 }{2\pi \omega}
\int_{E_-}^{E_+} d E_1E_1 E_2 \frac{1}{2} (\cdots)
\,, \hspace{0.5cm}
E_2 = \epsilon_{eg} - \omega - E_1
\,.
\end{eqnarray}
The angle factor constraint $|\cos \theta| \leq 1$
places a constraint on the range of neutrino energy integration,
\begin{eqnarray}
&&
E_{\pm} =
\frac{1}{2} \left( (\epsilon_{eg} - \omega) (1 +
\frac{m_i^2 - m_j^2}{\epsilon_{eg}(\epsilon_{eg} - 2\omega)} )
\pm \omega \Delta_{ij}(\omega)
\right)
\,,
\\ &&
\Delta_{ij}(\omega) 
= \left\{
\left(1 - \frac{ (m_i + m_j)^2}{\epsilon_{eg} (\epsilon_{eg} -2\omega) } \right)
\left(1 - \frac{ (m_i - m_j)^2}{\epsilon_{eg} (\epsilon_{eg} -2\omega) } \right)
\right\}^{1/2}
\,.
\end{eqnarray}

We record for completeness all four important integrals over the neutrino pair momenta:
\begin{eqnarray}
&&
\int d{\cal P}_{\nu} \frac{ 1}{ E_1E_2} = \frac{\Delta_{12}(\omega)}{2\pi} 
\equiv D_{12}(\omega)
\,,
\\ &&
\hspace*{-1cm}
\int d{\cal P}_{\nu} 1 = \frac{\Delta_{12}(\omega)}{2\pi}
\left(
\frac{1}{4} (\epsilon_{eg} - \omega)^2 - \frac{\omega^2 }{12}
 +\frac{\omega^2 (m_1^2 + m_2^2)}{6\epsilon_{eg}  (\epsilon_{eg} -2 \omega) }
- \frac{\omega^2 (m_1^2 - m_2^2)^2}{12 \epsilon_{eg}^2  (\epsilon_{eg} -2 \omega)^2} 
- \frac{ (\epsilon_{eg} - \omega)^2 (m_1^2 - m_2^2)^2}{2 \epsilon_{eg}^2  (\epsilon_{eg} -2 \omega)^2} 
\right)
\equiv C_{12}(\omega)
\,,
\nonumber \\ &&
\\ &&
\hspace*{-1cm}
\int d{\cal P}_{\nu} ( \frac{\vec{p}_1}{ E_1} + \frac{\vec{p}_2}{ E_2}) = 
- \frac{\Delta_{12}(\omega)}{4\pi}\vec{k}
\left(
\epsilon_{eg} -\frac{4}{3} \omega
 +\frac{2 (\epsilon_{eg} - \omega) (m_1^2 + m_2^2)}{3\epsilon_{eg}  
(\epsilon_{eg} -2 \omega) }
-\frac{4}{3} \frac{ (\epsilon_{eg} - \omega) (m_1^2 - m_2^2)^2}{ \epsilon_{eg}^2  (\epsilon_{eg} -2 \omega)^2} 
\right)
\equiv \vec{k}  \frac{J_{12}(\omega)}{\omega}
\,,
\label{vector integral}
\\ &&
\int d{\cal P}_{\nu} \frac{p_1^i  p_2^j +p_1^j p_2^i  }{2 E_1E_2} =
\frac{1}{2 } ( \delta_{ij} - \frac{k^i k^j}{\omega^2 } ) A_{12}(\omega)
 + \frac{1}{2 \omega^2 } (3\frac{k^i k^j}{\omega^2 } - \delta_{ij})
B_{12}(\omega)
 \,,
\\ &&
A_{12}(\omega) =\int d{\cal P}_{\nu} \frac{\vec{p}_1\cdot \vec{p}_2}{E_1 E_2}
\nonumber \\ &&
\hspace*{-1cm}
=  \frac{\Delta_{12}(\omega)}{2\pi} 
\left( 
- \frac{1}{4}  (\epsilon_{eg} - \omega)^2
+ \frac{5}{12} \omega^2 +  \frac{1}{2} (m_1^2 + m_2^2 )
 +\frac{\omega^2 (m_1^2 + m_2^2)}{6\epsilon_{eg}  (\epsilon_{eg} -2 \omega) }
- \frac{1}{12}  \frac{(m_1^2 - m_2^2)^2}{\epsilon_{eg}^2  
(\epsilon_{eg} -2 \omega)^2 }  
(\omega^2 + 3  (\epsilon_{eg} - \omega)^2\,)  
\right)
\,,
\\ &&
B_{12}(\omega) = \int d{\cal P}_{\nu} \frac{ \vec{k} \cdot\vec{p}_1 \vec{k} \cdot \vec{p}_2}{E_1 E_2}
= - \frac{\Delta_{12}(\omega)}{2\pi} 
\frac{\omega^2}{12} (\epsilon_{eg}^2 - 2 \omega\epsilon_{eg} - 2\omega^2 )
\,.
\end{eqnarray}

\vspace{0.5cm}
{\bf Appendix B: Magnetic factors}

\vspace{0.3cm}
It is important to clarify the magnetic field dependence
of PV observables, since this should help much to identify
RENP events in actual experiments.
In two types of diagrams of Fig(\ref {core renp pcpv 2})
and Fig(\ref {core renp pcpv 1}) the magnetic field dependence
is in atomic matrix elements of the form,
\begin{eqnarray}
&&
N_{PO, \pm}^{M} = 
\sum_{M_J} \langle ^1S_0| Y_{1,\pm 1} \widetilde{|^1P_1,M_J \rangle}
\widetilde{\langle ^3P_1, M_J |}
 S_z 
\widetilde{| ^3P_2, M \rangle}
\,,
\\ &&
N_{PE, \pm}^{M} = 
\sum_{M_J} \langle ^1S_0| Y_{1,\pm 1} \widetilde{|^1P_1,M_J \rangle}
\widetilde{\langle ^3P_1, M_J |}
\tilde{S}_z 
\widetilde{| ^3P_2, M \rangle}
\,,
\end{eqnarray}
where $\widetilde{|J, M \rangle} = e^{-i\theta_m J_y} | J, M\rangle$
is the  rotated state of a magnetic state, as
described in the text.
We need these functions for two
circularly polarized trigger of $\pm 1$ for E1 emission
as distinguished by the spherical harmonics $Y_{1,\pm 1}$.
Difference in two cases is in the spin component, either along
the fixed trigger axis in the PO case
 or along the magnetic field in the PE case.

PE case is easier to work out, since 
\begin{eqnarray}
&&
\widetilde{\langle ^3P_1, M_J | }\tilde{S}_z
\tilde{| ^3P_2, M \rangle} =
\langle ^3P_1, M_J| S_z | ^3P_2, M \rangle
\,.
\end{eqnarray}
The result is given using 3j symbols,
\begin{eqnarray}
&&
N_{PE, \pm}^{M}(x) =  - \sqrt{\frac{5}{2}} (-1)^{1-M}
\left(
\begin{array}{ccc}
1 & 1 & 2\\
-M & 0 & M
\end{array}
\right)
d^1_{M, \mp 1}(x)
\,.
\end{eqnarray}
More explicitly,
\begin{eqnarray}
&&
N_{PE, \pm}^{\mp 1}(x) = \frac{1}{12\sqrt{2}} (1 \pm \cos x )
\,, \hspace{0.5cm}
N_{PE, 0}^{\pm 1}(x) = \pm \frac{1}{6\sqrt{3}} \sin x
\,.
\end{eqnarray}
This gives $-W_{1, \pm}^{M}(x)$ in the text.

On the other hand, PO magnetic factors are written 
in terms of the product of three Wigner d-functions, and 
the final result is summarized by
\begin{eqnarray}
&&
\hspace*{-1cm}
N_{\pm 1, z}^{M} = 
- \sqrt{\frac{5}{2}} \sum_{|M_J, M_1| \leq 1}(-1)^{1-M_1}
\left(
\begin{array}{ccc}
1 & 1 & 2\\
-M_1 & 0 & M_1
\end{array}
\right)
d^1_{M_J, \mp 1} d^1_{M_J, M_1} d^2_{M, M_1}
= - \frac{1}{18}\sqrt{\frac{5}{2}} W_{2, \pm}^M
\,.
\end{eqnarray}
The final function is the one in the text.
Explicit forms are worked out:
\begin{eqnarray}
&&
W_{2, \mp}^{\pm} = \frac{1}{4} (\cos x + \cos (2x)\,)
\,, \hspace{0.5cm}
W_{2, \pm}^{\pm} = \frac{1}{4} (\cos x - \cos (2x)\,)
\,, \hspace{0.5cm}
W_{2, 0}^{\pm} = \pm \frac{1}{4}\sqrt{\frac{3}{2}}
\sin (2x)
\,.
\end{eqnarray}

On the other hand, magnetic factors of
PE amplitudes are given by $(W_{1, \pm}^M)^2$ for PE
and $W_{1, \pm}^M W_{2, \pm}^M $ for PO.
Their explicit forms are
\begin{eqnarray}
&&
{\rm PE \; squared \; amplitudes}; \;
(Mh) = (1,1), (-1, -1); \hspace{0.3cm}
\frac{1}{10} \cos^4 \frac{x}{2}
\,,
\\ &&
\hspace*{0.5cm}
(Mh) =
(1,-1), (-1,1) ; \hspace{0.3cm}
\frac{1}{10} \sin^4 \frac{x}{2}
\,, 
\\ &&
\hspace*{0.5cm}
(Mh) =
(0, \pm 1) ; \hspace{0.3cm}
 \frac{1}{15}\sin^2  x
\\ &&
{\rm PO \; squared \; amplitudes}; \;
(Mh) = (1,1), (-1, -1); \hspace{0.3cm}
\frac{1}{4} \sin^4 \frac{x}{2}(1+ 2 \cos x)^2
\,,
\\ &&
\hspace*{0.5cm}
(Mh) =
(1,-1), (-1,1) ; \hspace{0.3cm}
\frac{1}{4} \cos^4 \frac{x}{2}(1 - 2 \cos x)^2
\\ &&
\hspace*{0.5cm}
(Mh) =
(0, \pm 1) ; \hspace{0.3cm}
 \frac{3}{32}\sin^2  (2x)
\,.
\end{eqnarray}

Multiplying PO and PE amplitudes, one obtains
PV observables. The magnetic factor for PV observable thus derived is 
given by
\begin{eqnarray}
&&
{\rm PV \; observables}; \;
(Mh) = (1,1), (-1, -1); \hspace{0.3cm}
\frac{1}{2\sqrt{10}} \cos^4 \frac{x}{2} (1 - 2\cos x)
\,,
\\ &&
\hspace*{0.5cm}
(Mh) =
(1,-1), (-1,1) ; \hspace{0.3cm}
-\frac{1}{2\sqrt{10}} \sin^4 \frac{x}{2} (1 + 2\cos x)
\,, 
\\ &&
\hspace*{0.5cm}
(Mh) =
(0, \pm 1) ; \hspace{0.3cm}
\frac{1}{2\sqrt{10}}  \cos x \sin^2 x
\,.
\end{eqnarray}

\vspace{0.5cm}
{\bf Acknowledgements}
\hspace{0.2cm}
This research was partially supported by Grant-in-Aid for Scientific
Research on Innovative Areas "Extreme quantum world opened up by atoms"
(21104002)
from the Ministry of Education, Culture, Sports, Science, and Technology.


\begin{thebibliography}{99}
\bibitem{bouchiat 2}
M.A. Bouchiat and C. Bouchiat,
{\it J. Phys. (Paris)}{\bf 35}, 899 (1974);
{\it ibid.} {\bf 36},493 (1975).

\bibitem{bouchiat exp}
M.A. Bouchiat et al,
{\it Phys. Lett.}{\bf134B}, 463(1984),
and references therein.


\bibitem{commins}
P.S. Drell and E.D. Commins,
{\it Phys. Rev.}{\bf A 32}, 2196(1985),
and references therein.



\bibitem{wieman}
M.C. Noecker, B.P. Materson, and C.E. Wieman, 
{\it Phys. Rev. Lett.}{\bf 61}, 310 (1988),
and references therein.


\bibitem{nu oscillation data}
G. L. Fogli, E. Lisi, A. Marrone, D. Montanino, A. Palazzo, and A. M. Rotunno,
{\it Phys. Rev.} {\bf D 86}, 013012 (2012) [10 pages].

M. C. Gonzalez-Garcia, Michele Maltoni, Jordi Salvado, Thomas Schwetz,
{\it Journal of High Energy Physics} {\bf December 2012}, 123.

D. V. Forero, M. Toacutertola, and J. W. F. Valle,
{\it Phys. Rev.}{\bf D 86}, 073012 (2012) [8 pages].



\bibitem{tritium}
G. Drexlin, V. Hannen, S. Mertens, and C. Weinheimer, {\it Current Direct
Neutrino Mass Experiments},

Advances in High Energy Physics Volume 2013 (2013)Article ID 293986.


\bibitem{nu0 beta}
A. Gando et al,
{\it Phys. Rev. Lett.}{\bf 110}, 062502 (2013), and
arXiv:1201.4664v2[hep-ex] (2012).


M.Auger et al,
 {\it Phys. Rev. Lett.}{\bf 109}, 032505 (2012).





\bibitem{my-prd-07}
M. Yoshimura, {\it Phys. Rev.}{\bf D75}.
113007 (2007).

\bibitem{ptep overview}
A. Fukumi et al.,
{\it Progr. Theor. Exp. Phys.}{\bf 2012, 04D002};
arXiv1211.4904v1[hep-ph](2012).

\bibitem{ysu-pv-13}
M. Yoshimura, N. Sasao and S. Uetake,
{\it Parity violation in radiative emission of neutrino
pair from metastable states of heavy alkaline earth atoms},
arXiv 1312.6758 [hep-ph](2013).


\bibitem{b-assited clock transition}
A.V. Taichenachev et al,
{\it Phys. Rev. Lett.}{\bf 96}, 083001(2006);
Z.W. Barber et al,
{\it Phys. Rev. Lett.}{\bf 96}, 083002(2006).

\bibitem{ys-13}
M. Yoshimura and N. Sasao,
{\it Radiative emission of neutrino pair from
nucleus and inner core electrons in heavy atoms},
arXiv:1310.6472v1 [hep-ph](2013), and {\it Phys.Rev.D}
in press.

\bibitem{yst pra}
M. Yoshimura, N. Sasao, and M. Tanaka,
{\it Phys. Rev}
{\bf A86},013812(2012),
and
{\it Dynamics of paired superradiance},
arXiv:1203.5394[quan-ph] (2012).

\bibitem{dpsty-plb}
M. Yoshimura,
{\it  Phys. Lett.}{\bf B699},123(2011).

D.N. Dinh, S. Petcov, N. Sasao, M. Tanaka,
and M. Yoshimura,
{\it  Phys. Lett.}{\bf B719},154(2012), and  
arXiv1209.4808v1[hep-ph]. 

M. Tashiro et al,
{\it Progr. Theor, Exp. Phys.}, in press (2014).



\bibitem{atomic physics}
B.H. Bransden and C.J. Joachain, 
{\it Physics of Atoms and Molecules},
2nd edition, Prentice Hall(2003).


\bibitem{condon-shortley}
E.U.  Condon and G.H. Shortley,
{\it The Theory of Atomic Spectra},
Cambridge University Press (1951).

\bibitem{standard notation}
The notation used in atomic physics community
is different:
instead of $^{\pm}P_1$ here $^1\!P_1 (^+P_1) $ and $ ^3P_1 (^-P_1)$ are used,
along with $^1P_1^{(0)}, ^3\!P_1^{(0)}$ for our $^1P_1\,, ^3\!P_1$.
Our notation makes it more evident effect of
the intermediate coupling scheme using the $LS$ coupling basis.
Another minor difference is that our
$-\theta$ corresponds to the conventional $\theta$.


\bibitem{yfsy-10}
Results of the following paper by some of
us,
M. Yoshimura, A. Fukumi, N. Sasao, and
T. Yamaguchi
{\it Progr. Theor. Phys.}{\bf 123},523(2010),
contain effects linear in the applied
static Stark field, hence
the main part of its results reflects
the instrumental PV asymmetry rather than
the intrinsic PV asymmety of fundamental theory.



\bibitem{eta in ptep overview}
In \cite{ptep overview} a result for
numerical simulation of $\eta_{\omega}(t)$
is presented for pH$_2$ molecule target
(strong source of paired super-radiance (PSR)
of E1 $\times$ E1 transition, and  see
Fig 14 of this reference for time dependence).
Its time dependence is complicated:
a fast rise in $O(2$ ns), then a plateau region
of magnitude $O(10^{-2} \sim 10^{-3})$ of duration of
several nano-seconds, finally gradual decrease
ending around $10^{-6}$ at $\sim$ 12 ns (end time of calculation).
For RENP rate calculations,
numerical simulations based on the master
equation given in \cite{ptep overview} should be
performed for weaker PSR process of 
specific targets considered,
which is expected to give different time
profile and larger values of $\eta_{\omega}(t)$.



\bibitem{rose}
M.E. Rose,
{\it Elementary Theory of Angular Momentum},
Dover (1957).


\end{thebibliography}
\end{document}